\documentclass[twocolumn,showpacs,aps,prl,superscriptaddress]{revtex4}

\usepackage{xspace}
\usepackage{graphicx}
\usepackage{dcolumn}
\usepackage{amsmath}
\usepackage{epsfig}
\usepackage{multirow}

\input babarsym.tex

\newcommand{\lumi}             {\ensuremath{339\invfb}\xspace}

\newcommand{\Mnu}              {\ensuremath{m_{\nu}^2}\xspace}

\newcommand{\logmisspt}        {\ensuremath{-\ln (2\times p^T_{\mathrm{miss}}/\sqrt{s})}\xspace}

\newcommand{\tagmass}          {\ensuremath{m_{\mathrm{tag}}}\xspace}
\newcommand{\tagegam}          {\ensuremath{\Sigma E^{CM}_{\g}}\xspace}

\newcommand{\Egam}             {\ensuremath{E_{\g}}\xspace}

\def\DeltaE                    {\ensuremath{\Delta E}\xspace}

\newcommand{\eett}             {\ensuremath{e^+e^- \to \tautau}\xspace}

\def\eff                       {\ensuremath{\varepsilon}\xspace}

\def\tenseven                  {\ensuremath{\times 10^{-7}}\xspace}

\def\pzero                     {\ensuremath{P^0}\xspace}
\def\tautolpz                  {\ensuremath{\tau^\pm \to \ell^\pm \pzero}\xspace}
\def\tautoepz                  {\ensuremath{\tau^\pm \to \electron^\pm \pzero}\xspace}

\def\tautolpee                 {\ensuremath{\tau^\pm \to \ell^\pm\piz, \ell^\pm\eta, \ell^\pm\etapr}}

\def\tautoepiz                 {\ensuremath{\tau^\pm \to \electron^\pm \piz}\xspace}
\def\tautompiz                 {\ensuremath{\tau^\pm \to \mu^\pm \piz}\xspace}

\def\tautoeeta   {\ensuremath{\tau^\pm \to \electron^\pm \eta}\xspace}
\def\tautometa   {\ensuremath{\tau^\pm \to \mu^\pm \eta}\xspace}

\def\tautoeetap   {\ensuremath{\tau^\pm \to \electron^\pm \etapr}\xspace}
\def\tautometap   {\ensuremath{\tau^\pm \to \mu^\pm \etapr}\xspace}

\def\ptogg   {\ensuremath{\piz \to \gaga}\xspace}
\def\etogg   {\ensuremath{\eta \to \gaga}\xspace}
\def\etoppp  {\ensuremath{\eta \to \pi^+\pi^-\piz}\xspace}
\def\eptoppe {\ensuremath{\etapr \to \pi^+\pi^-\eta}\xspace}
\def\eptorg {\ensuremath{\etapr \to \rho^0\gamma}\xspace}
\def\rtopp  {\ensuremath{\rho^0\to\pi^+\pi^-}\xspace}

\newcommand{\tautoelg}  {\ensuremath{\tau \to e \nunub\gamma}\xspace}
\newcommand{\tautorho}  {\ensuremath{\tau \to \rho\nu}\xspace}

\newcommand{\roots}        {\ensuremath{\sqrt{s}}\xspace}

\def\kk         {\mbox{\tt KK2F}\xspace}
\def\tauola     {\mbox{\tt TAUOLA}\xspace}
\def\photos     {\mbox{\tt PHOTOS}\xspace}
\def\koralb     {\mbox{\tt KORALB}\xspace}
\def\evtgen     {\mbox{\tt EVTGEN}\xspace}
\def\jetset     {\mbox{\tt JETSET}\xspace}
\def\geant      {\mbox{\tt GEANT4}\xspace}

\newcommand{\gevccgevcc}{\ensuremath{{\mathrm{\,Ge\kern -0.1em V^2\!/}c^4}}\xspace}
\newcommand{\evcc}{\ensuremath{{\mathrm{\,e\kern -0.1em V\!/}c^2}}\xspace}
\newcommand{\CM} {\mbox{CM}\xspace}

\newcommand{\BABARPubYear}     {06}
\newcommand{\BABARPubNumber}  {061}
\newcommand{\SLACPubNumber} {12170}
\newcommand{\LANLNumber}  {0610067}

\def\figurebox#1#2#3{%
    \def\arg{#3}%
    \ifx\arg\empty
    {\hfill\vbox{\hsize#2\hrule\hbox to #2{\vrule\hfill\vbox to #1{\hsize#2\vfill}\vrule}\hrule}\hfill}%
    \else
    {\hfill\epsfbox{#3}\hfill}%
    \fi}

\begin{document}

\preprint{\babar-PUB-\BABARPubYear/\BABARPubNumber} 
\preprint{SLAC-PUB-\SLACPubNumber} 

\begin{flushleft}
\babar-PUB-\BABARPubYear/\BABARPubNumber\\
SLAC-PUB-\SLACPubNumber\\
hep-ex/\LANLNumber\\[10mm]    
\end{flushleft}

\title{
{\large \bf \boldmath Search for Lepton Flavor Violating Decays \tautolpee}
}

%Input author list file
%
%% author list as of 01-Sep-2006 (603 authors)
%
\author{B.~Aubert}
\author{M.~Bona}
\author{D.~Boutigny}
\author{F.~Couderc}
\author{Y.~Karyotakis}
\author{J.~P.~Lees}
\author{V.~Poireau}
\author{V.~Tisserand}
\author{A.~Zghiche}
\affiliation{Laboratoire de Physique des Particules, IN2P3/CNRS et Universit\'e de Savoie, F-74941 Annecy-Le-Vieux, France }
\author{E.~Grauges}
\affiliation{Universitat de Barcelona, Facultat de Fisica, Departament ECM, E-08028 Barcelona, Spain }
% ...
\collaboration{The \babar\ Collaboration}
\noaffiliation

\date{\today}% It is always \today, today, but you may specify any date with \date.

\begin{abstract}
A search for lepton flavor violating decays of the \mtau lepton to a lighter mass lepton and a pseudoscalar meson
has been performed using \lumi of \epem annihilation data collected at a center-of-mass energy near 10.58\gev 
by the \babar\ detector at the SLAC PEP-II storage ring.
No evidence of signal has been found,
and upper limits on the branching fractions are set at $10^{-7}$ level.
\end{abstract}

\pacs{13.35.Dx, 14.60.Fg, 11.30.Hv}

\maketitle

%%
%% --------- Introduction ----------------
%%

The recent discovery of large neutrino mixing~\cite{NuOsc} suggests that lepton flavor violation (LFV) occurs.
Charged LFV decays have not yet been observed, 
although they have long been identified as unambiguous signature of new physics.
Neutrinoless decays like \tautolpz, where $\ell = e, \mu$ and 
$\pzero = \piz, \eta, \etapr$, are likely candidates for LFV~\cite{SusyLFV,Sher:2002ew},
which could be induced by potentially large mixing between the supersymmetric partners of the leptons
and is further enhanced by color factors associated with these semi-leptonic decays.
Some models with heavy Dirac neutrinos~\cite{Gonzalez-Garcia:1991be,Ilakovac:1999md},
two Higgs doublet models, R-parity violating supersymmetric models,
and flavor changing $Z^\prime$ models with non-universal couplings~\cite{Li:2005rr}
allow for observable parameter space of new physics~\cite{Black:2002wh},
while respecting the existing experimental bounds~\cite{belle}.

%% --------- Data set and Detector ----------------
%%
The results presented here use an integrated luminosity \L = \lumi
collected at a center-of-mass (\CM) energy, \roots, near 10.58\gev 
by the \babar\ detector at the SLAC \pep2 \epem asymmetric-energy storage ring.
Details of the \babar\ detector are described elsewhere~\cite{detector}.

%%
%% --------- Selection ----------------
%%
The signature of the signal process is the presence of an $\ell\pzero$ pair 
having an invariant mass consistent with $m_\tau$ = 1.777\gevcc~\cite{bes}
and a total energy equal to \roots/2 in the \CM frame, 
along with other particles in \eett events having properties consistent with a \mtau lepton decay.
Two neutral decay modes (\ptogg and \etogg) and 
three charged decay modes [\etoppp (\ptogg), \eptoppe (\etogg), and \eptorg] are reconstructed.
Signal events are simulated with the \kk~\cite{kk} Monte Carlo (MC) program,
where the \tautolpz decays according to two body phase space,
while the other $\tau$ decays according to measured branching fractions~\cite{Yao:2006px}
simulated with \tauola~\cite{tauola}.
$\mu^+\mu^-$ and \tautau background processes are generated using \kk and \tauola,
and \qqbar processes are generated using \evtgen~\cite{Lange:2001uf} and \jetset~\cite{Sjostrand:1995iq}.
Radiative corrections are simulated using \photos~\cite{Golonka:2005pn}.
The detector response is simulated with \geant~\cite{geant}.
The MC events are used for the optimization and systematic studies
of the signal efficiencies, and for determination of the background
shapes. Estimates of the rates for the backgrounds are derived
directly from the data.

Events with two or four well reconstructed tracks and zero total charge are selected.
Tracks are rejected if they are consistent with coming from photon conversions.
An event is divided into two hemispheres (``signal''- and ``tag''- sides) in the CM frame 
by a plane perpendicular to the thrust axis~\cite{thrust},
calculated using all observed particles.

The signal-side hemisphere is required to contain one or three tracks and two photon candidates
with energy \Egam $>$ 50\mev for the \ptogg, \etoppp (\ptogg) and \eptoppe (\etogg) channels,
and \Egam $>$ 100\mev for the \etogg channel.
For the \eptorg channel, the single photon candidate is required to have \Egam $>$ 100\mev.
Events with additional photon candidates in the signal hemisphere with \Egam $>$ 100\mev are rejected.

The \pzero candidates are reconstructed in the following mass windows:
$m(\ptogg)$   $\in$ [0.115, 0.150] \gevcc,
$m(\etogg)$   $\in$ [0.515, 0.565] \gevcc,
$m(\etoppp)$  $\in$ [0.537, 0.558] \gevcc,
$m(\eptoppe)$ $\in$ [0.950, 0.965] \gevcc,
$m(\eptorg)$  $\in$ [0.940, 0.970] \gevcc, and
$m(\rtopp)$   $\in$ [0.600, 0.900] \gevcc.
To reduce combinatorial backgrounds, the momentum of \pzero is required to satisfy:
$p_{\piz}$ $>$ 0.5\gevc for \tautoepiz,
$p_{\piz}$ $>$ 1.5\gevc for \tautompiz, 
$p_{\eta}$ $>$ 1.0\gevc for \tautoeeta (\etogg),
$p_{\eta}$ $>$ 1.4\gevc for \tautometa (\etogg),
$p_{\eta}$ $>$ 1.2\gevc for \tautometa (\etoppp, \ptogg) and
$p_{\etapr}$ $>$ 1.3\gevc for \tautometap (\eptorg) decays.

The track unassociated with any of the $\pzero$ daughters 
is required to have a momentum $>$ 0.5\gevc and 
is identified as an electron or muon, but not as a kaon, 
using standard \babar\ particle identification techniques~\cite{Aubert:2002rg}.
In the case of a charged $\pzero$ decay, 
this criteria is applied on the track that combines with the one having opposite-sign charge and 
the photon candidate(s) to give an invariant mass farthest from the nominal $\pzero$ mass~\cite{Yao:2006px}.
This provides the correct pairing for $>99.7\%$ of selected signal MC events 
after particle-identification requirements.

The origin of the photon(s) is assigned to the point of closest approach of the lepton track 
to the $\epem$ collision axis for neutral $\pzero$ 
decays, or to the common vertex in the signal-side hemisphere for the charged $\pzero$ decays.
The $\pzero$ momentum is kinematically fitted with its respective mass constraints,
and combined with the lepton track to form the signal \mtau candidate.
An event is accepted 
based upon the closeness of the signal \mtau candidate to $m_\tau$.

\begin{figure*}[!htbp]
\caption{Selected data (dots) and 68\% of signal MC events (shaded region) inside the GSB region, and the $\pm2\sigma$ signal box.}
\resizebox{\textwidth}{.3\textheight}{%
\includegraphics{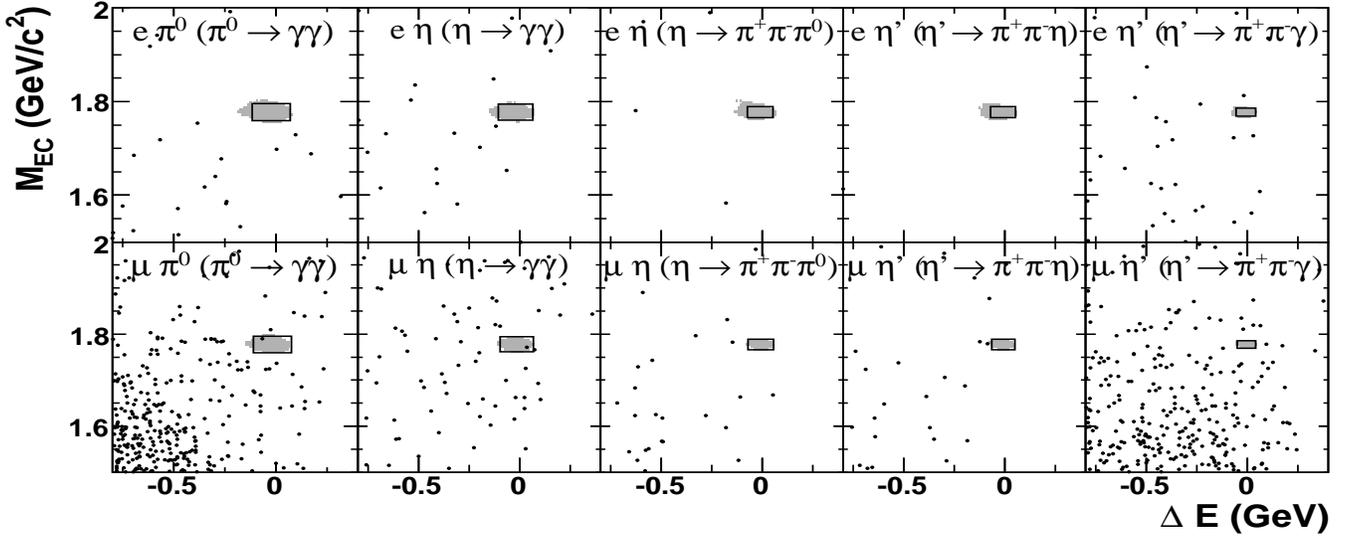}}
\label{fig1}
\end{figure*}

Signal decays are identified by two kinematic variables: 
the beam-energy constrained $\tau$ mass \mec and $\DeltaE \equiv E_{\ell} + E_{\pzero} - \roots/2$, 
where $E_\ell$ and $E_{\pzero}$ are the respective energies in the \CM frame.
These two variables are independent apart from small correlations 
arising from initial and final state radiation.
For signal events, the reconstructed peak positions of the \mec distribution agree very well with $m_\tau$,
while those of \DeltaE vary between $-5$ to $-23 \mev$.
The shift from zero in the \DeltaE peak comes from mis-calibration of the measured photon energy. 
The resolutions of the \mec and \DeltaE distributions for the signal events 
are presented in Table~\ref{table1}.
Events in the data within a $\pm$3$\sigma$ rectangular box centered around the signal MC peak positions
are excluded until all optimization and systematic studies of the selection criteria have been completed.
The selection is optimized to yield the smallest expected upper limit~\cite{Feldman:1997qc}
in a background-only hypothesis for observing events inside the $\pm2\sigma$ rectangular signal box
around the signal MC peak positions shown in Fig.~\ref{fig1}.

The dominant backgrounds are from \tautoelg or \tautorho decays in \tautau events,
with additional contributions from Bhabha, di-muon and $\qqbar$ processes.
The backgrounds are higher for searches with muons, due to misidentification 
of a $\pi$ track as a $\mu$ candidate. Another source of background is the 
mis-reconstruction of $\eta$ and $\piz$ candidates.

Non-$\tau$ backgrounds with radiation along the beam directions are suppressed by requiring
the polar angle of the missing momentum to lie between $-0.76$ and $0.92$.
The total \CM momentum of all tracks and photon candidates on the tag-side 
is required to be less than 4.75\gevc.

A tag-side hemisphere containing a single track is classified as $e$-tag, $\mu$-tag or  $h$-tag
if the track is exclusively identified as an electron, muon, or neither, respectively.
For these tags, the total neutral \CM energy in the hemisphere \tagegam is required to be less than 0.2\gev,
and the invariant mass \tagmass, calculated using all observed charged and neutral particles, to be less than 0.4\gevcc. 
For \tautoepz channels, the data events in $e$-tag are used as a control sample to estimate the Bhabha background,
and are not included in the final selection.
If the track is neither an electron nor a muon, \tagegam $>$ 0.2\gev and \tagmass $\in$ [0.6, 1.3] \gevcc,
the event is classified as a $\rho$-tag. 
For searches of neutral $\pzero$ decay modes,
events with three tracks in the tag-side with \tagmass $\in$ [0.9, 1.6] \gevcc are also allowed.

Taking the direction of the tag-side $\tau$ to be opposite the signal candidate,
all tracks and photon candidates in the tag-side hemisphere are used to calculate the
invariant mass squared of the tag-side missing momentum (\Mnu).
To reduce non-$\tau$ backgrounds
for \tautoepiz, \tautoeeta (\etogg) searches,
$(\Mnu/1.8\gevccgevcc)\logmisspt/2.0$ is required to be less than unity~\cite{Aubert:2005wa},
where $p^{T}_{miss}$ is the component of the missing momentum transverse to the collision axis.
For the other searches, \logmisspt is required to be less than 2.5, 
except for \tautoeeta (\etoppp) and \tautometap (\eptoppe) searches, where very few events are expected.

To focus on selected signal-like events,
a Grand Side Band (GSB) is defined in the \mec vs. \DeltaE plane as:
\mec $\in$ $[1.5, 2.0]$ \gevcc and \DeltaE $\in$ $[-0.8, 0.4]$ \gev.
With electrons as the lepton track, 22, 18, 4, 1 and 30 events survive in the GSB 
for \ptogg, \etogg, \etoppp, \eptoppe and \eptorg channels, 
and 311, 69, 24, 24 and 285 events survive for the corresponding channels with muons as the lepton track,
as shown by dots in Fig.~\ref{fig1}.
Also shown are the shaded regions containing 68\% of the selected signal MC events inside the GSB.

\renewcommand{\multirowsetup}{\centering}
\newlength{\LL}\settowidth{\LL}{$8047$}
\begin{table*}[!htbp]
\begin{center}
\caption{The \mec and \DeltaE resolutions for the signal MC events,
the number of observed (obs.) and expected (exp.) events 
inside $\pm3\sigma$ to $\pm11\sigma$ boxes and $\pm2\sigma$ box,
the branching fractions (\BR), the efficiencies (\eff), and the 90\% C.L. upper limits (UL).}
\label{table1}
\begin{ruledtabular}
\renewcommand{\arraystretch}{1.10}
{\scalebox{0.99}{
\begin{tabular}{l|c|c|c|r@{$~~~$}|c|c|c|c|c|c}
Decay modes&$\sigma(\mec)$&$\sigma(\DeltaE)$
&\multicolumn{2}{c|}{$\pm3\sigma$ to $\pm11\sigma$ box}&\multicolumn{2}{c|}{$\pm2\sigma$ box}
&\BR&\eff&\multicolumn{2}{c}{UL (\tenseven)}\\\cline{2-11}
                      &\mevcc&\mev&obs.&\multicolumn{1}{c|}{exp.} &obs.&exp.      &(\%)          &(\%)         &obs.&exp.\\\hline
\tautoepiz (\ptogg)   &9.1&46.4& 4& 5.37$\pm$1.14 &0&0.17$\pm$0.04&98.80$\pm$0.03&2.83$\pm$0.25&1.3&1.4\\\hline\hline
\tautompiz (\ptogg)   &9.0&46.4&43&40.68$\pm$4.32 &1&1.33$\pm$0.15&98.80$\pm$0.03&4.75$\pm$0.37&1.1&1.1\\\hline\hline
\tautoeeta (\etogg)   &8.5&42.6& 4& 4.99$\pm$1.18 &0&0.20$\pm$0.05&39.38$\pm$0.26&3.59$\pm$0.24&2.5&2.8\\\hline
\tautoeeta (\etoppp)  &5.9&31.4& 0& 0.64$\pm$0.32 &0&0.02$\pm$0.01&22.43$\pm$0.40&3.17$\pm$0.32&5.4&5.5\\\hline
\tautoeeta              &&&&&0&0.22$\pm$0.05&\multicolumn{2}{c|}{\BR\eff = 2.12$\pm$0.20 (\%)}&1.6&1.9\\\hline\hline
\tautometa (\etogg)   &8.3&40.8&20&17.36$\pm$2.12 &1&0.67$\pm$0.08&39.38$\pm$0.26&7.03$\pm$0.53&1.9&1.6\\\hline
\tautometa (\etoppp)  &5.6&31.0& 3& 2.01$\pm$0.41 &0&0.08$\pm$0.02&22.43$\pm$0.40&3.67$\pm$0.32&4.5&4.8\\\hline
\tautometa              &&&&&1&0.75$\pm$0.08&\multicolumn{2}{c|}{\BR\eff = 3.59$\pm$0.41 (\%)}&1.5&1.3\\\hline\hline
\tautoeetap(\eptoppe) &5.9&31.0& 0& 0.14$\pm$0.14 &0&0.01$\pm$0.01&17.52$\pm$0.56&3.75$\pm$0.27&5.8&5.9\\\hline
\tautoeetap(\eptorg)  &4.4&24.3& 2& 2.97$\pm$0.54 &0&0.11$\pm$0.03&29.40$\pm$0.90&2.98$\pm$0.28&4.2&4.5\\\hline
\tautoeetap             &&&&&0&0.12$\pm$0.03&\multicolumn{2}{c|}{\BR\eff = 1.53$\pm$0.16 (\%)}&2.4&2.6\\\hline\hline
\tautometap(\eptoppe) &5.6&29.1& 1& 2.42$\pm$0.47 &0&0.07$\pm$0.02&17.52$\pm$0.56&5.87$\pm$0.46&3.6&3.8\\\hline
\tautometap(\eptorg)  &4.1&23.1&13&11.06$\pm$0.65 &0&0.42$\pm$0.03&29.40$\pm$0.90&3.90$\pm$0.46&2.7&3.7\\\hline
\tautometap             &&&&&0&0.49$\pm$0.04&\multicolumn{2}{c|}{\BR\eff = 2.18$\pm$0.26 (\%)}&1.4&2.0\\
\end{tabular}
}}
\end{ruledtabular}
\end{center}
\end{table*}

The number of expected background events in the signal box is extracted from
an unbinned maximum likelihood fit to the distributions of \mec and \DeltaE in data
inside the non-blinded parts of the GSB, using two-dimensional probability density functions (PDF)
for \epem, \mumu, \tautau and \qqbar backgrounds.
The kernel of the PDFs are estimated~\cite{keys} from the data control samples for \epem,
and respective MC events for the others.

The dominant contribution to the uncertainty in background estimation arises from the statistical precision 
on the selected data sample inside the non-blinded parts of the GSB, 
or from the variation of background components within $\pm$1$\sigma$ from their fitted values.
The observed and expected events from the fit to the data 
inside the $\pm3\sigma$ to $\pm11\sigma$ annular boxes and the signal boxes
are shown in Table~\ref{table1}, which confirm good modeling of the backgrounds in data and show no evidence of signal.

The largest systematic uncertainties in the signal reconstruction efficiency are due to the signal track momentum and 
the photon energy scale and resolution, estimated by varying the peak position and resolution of the \mec and \DeltaE distributions.
The errors associated with the modeling of each selection variable are estimated from the
relative change in signal efficiency when varying the selection criteria 
by the difference between the data and MC events in the mean of that variable.
Other sources of systematic uncertainties include those arising from trigger inefficiencies,
tracking and neutral energy reconstruction efficiencies, the signal lepton identification,
beam-energy scale and spread, luminosity estimation and 
$\epem\to\tautau$ cross-section $(\sigma_{\tau\tau}=0.89\pm0.02~\mathrm{nb})$~\cite{kkxsec}.
About 2.4 million MC events are used per channel, resulting in a negligible systematic uncertainty due to MC statistics.
Although the signal MC events have been modeled using a two body phase space model, 
the results obtained in this analysis are insensitive to this assumption 
as demonstrated by considering the two extreme cases of a $V-A$ and a $V+A$ form of interaction.

The upper limits for \tautolpz decays are calculated using 
$\BR^{90}_{UL}=N^{90}_{UL}/(2\L\sigma_{\tau\tau}\BR\eff)$,
where $N^{90}_{UL}$ is the 90\% confidence level (C.L.) upper limit 
on the number of signal events inside the signal box, 
\BR\ and \eff are the branching fraction~\cite{Yao:2006px} 
and reconstruction efficiency of the signal decay mode under consideration.
To obtain a combined upper limit with $\eta$ and $\etapr$ decays, 
the observed and expected background events
and the signal efficiencies are added using $\BR\eff = (\BR_1 \times \eff_1 + \BR_2 \times \eff_2)$,
where $\BR_1$, $\BR_2$ are the respective branching fractions
and $\eff_{1}$ and $\eff_{2}$ are the corresponding efficiencies.
This combination takes into account correlated uncertainties
from the track and neutral cluster reconstruction efficiency and the signal lepton identification.
The observed and the expected upper limits at 90\% C.L. are presented in Table~\ref{table1} 
including all contributions from systematic uncertainties~\cite{Cousins:1992qz,Barlow:2002bk}.
These limits present up to a factor of four improvement 
over the previously published results~\cite{belle},
except for \tautometa search, where the limit is similar.

\begin{figure}[!hbtp]
\caption{Excluded regions in $\tan\beta$ vs. $m_A$ plane (see text).}
\resizebox{.96\columnwidth}{.3\textheight}{%
\includegraphics{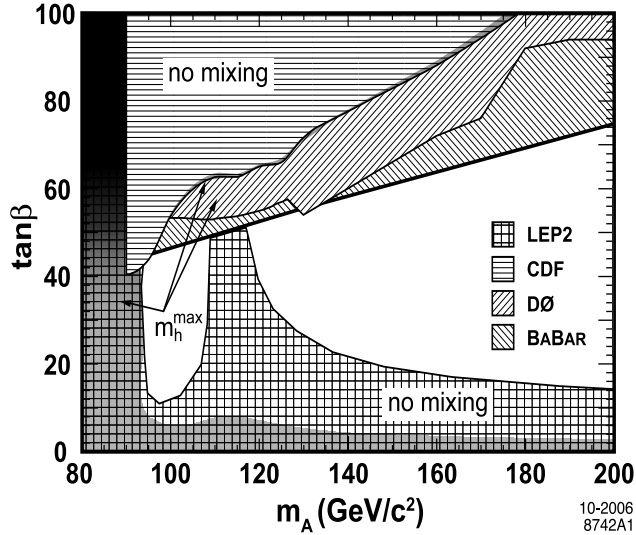}}
\label{fig2}
\end{figure}

Mixing between left-handed smuons and staus allows one to translate the \tautometa limit to an exclusion plot in
the $\tan\beta$ vs. $m_A$ plane~\cite{Sher:2002ew}, 
where $\tan\beta$ is the ratio of the vacuum expectation values 
of the two Higgs doublets and $m_A$ is the mass of the pseudoscalar Higgs.
The excluded regions at 95\% C.L. from this \tautometa search ($<1.9\tenseven$) 
with right-handed neutrino mass = 10$^{14}$\gevcc introduced via the seesaw mechanism
are shown in Fig.~\ref{fig2}. This result is competitive with those obtained 
from the direct searches for Higgs $\to$ $b\bar{b}$, $\tautau$ decays
by CDF~\cite{cdf} and D0~\cite{dzero}, and complementary to the region 
excluded by the LEP experiments with a top quark mass of 174.3 \gevcc~\cite{lephiggs}, 
for two common scenarios of stop-mixing benchmark models~\cite{Carena:1999xa}: 
$m_h^{\rm{max}}$ and no-mixing obtained with the Higgs mass parameter $\mu = -200 \gevcc$ 
shown by darker and lighter shaded regions respectively.

% Input the pubboard acknowledgements file
We are grateful for the excellent luminosity and machine conditions
provided by our \pep2\ colleagues, 
and for the substantial dedicated effort from
the computing organizations that support \babar.
The collaborating institutions wish to thank 
SLAC for its support and kind hospitality. 
This work is supported by
DOE
and NSF (USA),
NSERC (Canada),
IHEP (China),
CEA and
CNRS-IN2P3
(France),
BMBF and DFG
(Germany),
INFN (Italy),
FOM (The Netherlands),
NFR (Norway),
MIST (Russia),
MEC (Spain), and
PPARC (United Kingdom). 
Individuals have received support from the
Marie Curie EIF (European Union) and
the A.~P.~Sloan Foundation.

%%%%%%%%%%%%%%%%%%%%%%%%%%%%%%%%%%%%%%%%%%%%%%%%%%%%%%%%%%%%%%%%%%%%%%%%%%%%%%%%%%%%%%%%%%%%%%%

\end{document}